\documentclass{aastex}
\usepackage{spr-astr-addons}

\usepackage{url}
\usepackage{graphics}

\RequirePackage{color}

\def\vec#1{\mathbf{#1}}

\begin{document}

\title{Presupernova Structure of Massive Stars}
\slugcomment{HEDLA 2010, 8th International Confernce on High Energy Density Laboratory Astrophysics, California Institute of Technology}
\shorttitle{Presupernova Structure}
\shortauthors{Meakin \& Arnett}

\author{Casey A. Meakin}\author{Tuguldur Sukhbold}\author{W. David Arnett}
\affil{Steward Observatory, University of Arizona, Tucson, AZ, 85721, USA}
\email{casey.meakin@gmail.com}

\begin{abstract}
Issues concerning the structure and evolution of core collapse progenitor stars are discussed with an emphasis on interior evolution. We describe a program designed to investigate the transport and mixing processes associated with stellar turbulence, arguably the greatest source of uncertainty in progenitor structure, besides mass loss, at the time of core collapse.  An effort to use precision observations of stellar parameters to constrain theoretical modeling is also described.
\end{abstract}



\section{Introduction}

\par To first order, stellar evolution theory is a success: an evolutionary picture unifying the gross statistical properties of stars and stellar populations has been developed. However, a number of current astrophysical puzzles demand a more accurate theory of star evolution than currently exists.  Examples include (but are in no way restricted to): (1) understanding the initial conditions for explosions in massive stars, including core collapse supernovae and gamma ray bursts, (2) the formation of O-Ne-Mg cores and the possibility of accretion induced collapse (AIC), (3) the nucleosynthesis in AGB stars, including heavy element  (Z$\gtrsim$26) formation through the slow neutron capture process, or {\em s-process}, and (4) the internal structure of the Sun, and the {\em tachocline} in particular.  All of these research areas rely heavily on stellar evolution calculations and suffer from unacceptably large theoretical uncertainties, in most cases related to an inadequate treatment of turbulent mixing processes.

\section{The current framework for computing evolution}
\label{sec:framework}

\par {\em A 1D Model.} The basic framework for stellar evolution relies on the approximation of spherical symmetry. When rotation is strong enough to distort the star's shape from spherical symmetry the radial coordinate is replaced by an equipotential coordinate and quantities averaged over equipotential surfaces are evolved instead, thus retaining a one-dimensional treatment \citep{kippenhahn1970,endal1976}.

\par The hydrodynamics governing stellar interiors are grossly simplified from the outset by this reduction in dimensionality, and a quasi-static approximation is generally adopted.  Any additional, non-radial hydro- and magneto-dynamic processes (such as convection and internal waves) are then incorporated on top of this quasi-static, 1D evolution in the form of transport processes, which will be discussed further below.  The quasi-static equations for single star structure and evolution \citep[e.g.][]{kippenhahn1990} involve the 1D formulations of mass, momentum, and energy conservation together with the auxiliary equations

\begin{eqnarray}
\label{eq:transport}
\frac{\partial T}{\partial m} &=&  -\frac{G m T}{4\pi r^4 P}\nabla\\
\label{eq:composition-evolution}
\frac{\partial X_i}{\partial t} &=& \dot{X}_{i}^{\rm nuc} - \nabla\cdot\vec{J}_i
\end{eqnarray}

\noindent which describe the mode of energy transport (eq.~\ref{eq:transport}) and the compositional evolution of the stellar plasma (eq.~\ref{eq:composition-evolution}), where standard notation has been used and the temperature gradient is written in terms of $\nabla = d\ln T/d\ln P$ (not to be confused with the divergence operator in eq.~\ref{eq:composition-evolution}).  The compositional flux $\vec{J}_i$ is due to the cumulative effects of a variety of processes including turbulence, internal wave motions, and radiative levitation and elemental settling, while $\dot{X}_i^{\rm nuc}$ is the change in composition due to nuclear burning.

\par The time evolution of the star is driven by the nuclear transmutation of matter and its heating, energy losses at the surface, neutrino cooling in the interior, and the indelible mixing of composition and angular momentum.  The theory of stellar structure and evolution consists of trying to capture the dynamics of this complex system with fidelity.  Additional equations are required to incorporate the effects of angular momentum and magnetic fields, as well as mass loss from the surface and binary interactions, including tidal distortion and mass transfer. 

\par {\em Energy Transport and Convection.} But even when considering non- rotating, non- magnetic, single star evolution the theory remains incomplete.  Energy transport and thermal instabilities in the stellar interior, and their attendant mixing processes, remain some of the most pernicious outstanding issues. Energy is carried by radiation, thermal convection, or a combination of both. The theories underlying these different modes of energy transport determine the equilibrium temperature gradient $\nabla$ appearing in eq.~\ref{eq:transport}.  Radiation transport is simplified in the deep interior where photon mean free paths (m.f.p.s) are small (compared to any radial scale height in the star) and a diffusion approximation is accurate. In this case the temperature structure depends primarily on the opacity of the plasma $\kappa$ and one finds $\nabla\rightarrow\nabla_{\rm rad}  = 3\kappa LP/(16\pi a c G m T^4)$.  

\par When $\nabla_{\rm rad} > \nabla_{\rm ad} \equiv (d\ln T/d\ln P)_s$ the stellar plasma is unstable to thermal convection and a hydrodynamic flow ensues and modifies the underlying temperature stratification. In these regions a model for convective energy transport is required to calculate the temperature (or entropy) profile of the star. The almost universally adopted model is mixing length theory (MLT) which provides a functional dependence for $\nabla$ on the stellar luminosity, structure, and microphysical properties of the plasma (e.g., opacities, equation of state, nuclear burning rates) which is only slightly more complicated than the expression for $\nabla_{\rm rad}$ defined above.

\par Finally, what is arguably the most complex situation regarding energy transport is when the convective instability extends into the photosphere in which case the diffusion approximation breaks down and one needs to treat simultaneously thermal convection and a complex radiation transport problem.  In this case as well, the almost universally adopted approach is to use MLT and radiative diffusion in evolution models, despite the obvious physical shortcomings \citep[see][for a discussion of the solar case]{arnett2010}.  Three dimensional radiation-hydrodynamic simulations  \citep[e.g.,][]{nordlund1982,steffen1991,stein1998,robinson2003} are beginning to provide self-consistent solutions to this complex problem, facilitating detailed comparison between model and observation.  A stronger foundation for interpreting helioseismic data \citep[e.g.][]{stein2001,li2002} and inferring the abundances of various elemental constituents of the sun \citep[e.g.][]{asplund2005,caffau2008} are benefitting tremendously from this work: these almost brute force approaches to solar surface convection are offering sufficiently precise information about interior structure to shed light on deeper astrophysical conundrums including the formation scenario of the sun \citep{guzik2010} and, in combination with solar neutrino observations, standard model physics \citep{serenelli2009}.

\par {\em Other Instabilities and Mixing.}  Instabilities involving composition gradients \citep{langer1983,spruit1992,eggleton2006,mocack2010b}, differential rotation \citep{pinsonneault1989}, and magnetic fields \citep{spruit2002} are just as important as thermal convection with regards to interior mixing. This mixing is almost invariably treated as a diffusive process, taking the divergence of the compositional flux to be

\begin{equation}
\label{eq:diffusion}
-\nabla\cdot\vec{J}_i = \frac{\partial}{\partial m}\Big(\hat{D}\frac{\partial X_i}{\partial m}\Big).
\end{equation}

\noindent The Eulerian diffusion coefficient $D = ({\partial m}/{\partial r})^{-2}\hat{D}$ is most often formulated from a combination of characteristic length $l$, velocity $v$, and time scale $\tau$, computed from linear stability theory \citep{chandrasekhar1961,drazin2004}, with $D \sim l v$ or $D\sim l^2/\tau$. \citet{heger2000,heger2005} provides a fairly comprehensive presentation of stellar physics in this "diffusive" spirit. Caution is advised when using these results, however, since the development and mixing rates for these instabilities are not necessarily being represented correctly by this approach, particularly in light of the fact that the high Reynolds number characterizing these flows ensure that the subsequent development will be highly turbulent and in the non-linear regime.  See also \citet{bruenn2005} for a related discussion on the modern stellar evolution model and its shortcomings in core-collapse progenitor modeling.

\par Two big open question are: (1) Is a diffusive treatment for mixing (as in eq.~\ref{eq:diffusion}) appropriate? (2) If not, how shall we improve this situation?  Two processes have already been shown to differ dramatically from diffusion, including the flux of kinetic energy and the erosion of convective boundary layers due to convective overshoot. The kinetic energy flux is now known to carry a non-negligible fraction of the total luminosity of the star under many circumstances, contrary to MLT which takes this flux to be zero.  In the rare case that this contribution to the luminosity is included in a stellar model it is usually treated using a ``down gradient approximation'' with $F_K \propto -\nabla E_K$ \citep{kuhfuss1986}, which is in direct conflict with simulated flow as well as basic considerations of energy conservation \citep{meakin2010}.  Similarly, the evolution of convective boundaries are likely to evolve in a manner which is far from diffusive in nature, and behave instead as sharp entrainment interfaces, which will be discussed further in \S\ref{sec:future}.

\section{How well does the current theory work?}


\begin{figure}[t]
\mbox{\includegraphics[width=3.3in]{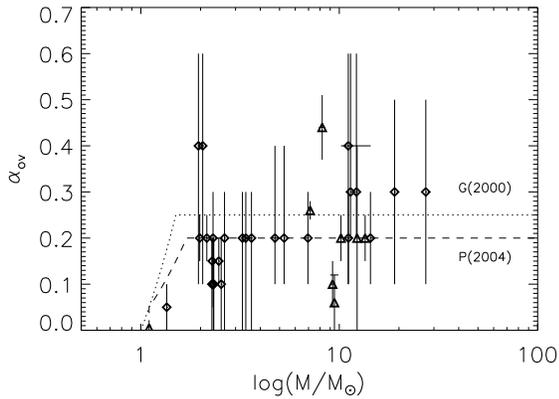}}
\caption{\label{fig:ov-mass} Overshoot parameter estimates from wide eclipsing binary (diamonds) and asteroseismological data (triangles). The values adopted by \citet{girardi2000} and \citet{pietrinferni2004} for populations synthesis are labeled G2000 and P2004.  The mass uncertainties for the eclipsing binary data is negligible on the log(M) scale used. Overshoot parameters inferred from asteroseismic data are from \citet{soriano2010}, \citet{dimauro2003}, \citet{suarez2009}, \citet{briquet2007}, \citet{dupret2004}, \citet{aerts2006}, \citet{desmet2009}, and\citet{mazumdar2006}; and the values from eclipsing binary data are from \citet{claret2007}.}
\end{figure}

\begin{figure}[t]
\mbox{\includegraphics[width=3.3in]{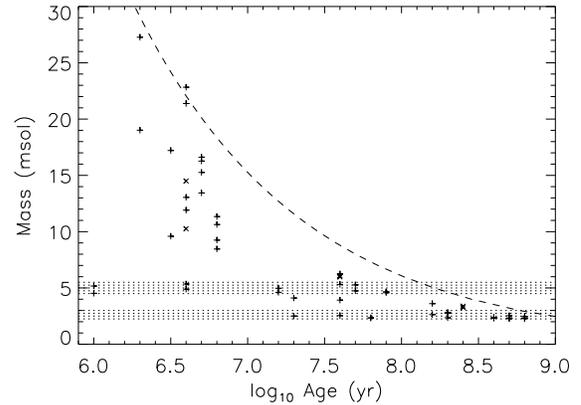}}
  \caption{\label{fig:age-mass} The ages and masses of wide eclipsing binary star data having precision observational parameters from \citet{torres2010}.  These are all upper main sequence stars with convective cores and radiative envelopes and are therefore some of the simplest stellar objects to model.  The dashed curve represents an approximate main sequence turn off age as a function of initial mass.}
\end{figure}

\par {\em Quantifying Uncertainty.} While the uncertainty inherent in the stellar modeling framework is difficult to quantify, a formal lower bound can be assessed by comparing stellar parameters calculated in the following ways.  
(1) The results from a variety of evolution codes which have nominally the same  input physics and parameters can be compared. This exercise addresses the question: How accurately is the mathematical model being captured by the numerical implementation?
(2) The results from a single code using a range of plausible adjustable parameters and input physics can be compared.  This addresses the question: How sensitive is the stellar model to uncertainties in the input physics and parameters within the framework being used?

\par While some effort has been exerted to obtain consistency between different implementations of the same physical model \citep[e.g.][]{montalban2008,marconi2008}, code to code scatter persists even for the relatively simple case of main sequence evolution.   
A recent comparison between the inferred properties of massive O and B stars within 500 pc \citep{hohle2010} using three modern evolution codes \citep{schaller1992,bertelli1994,claret2004} indicate uncertainties at the $\sim$20 - 25 \% level.   Similar degrees of uncertainty were found when computing the mass limits associated with O-Ne-Mg core formation in intermediate mass stars \citep{poelarends2008}.

\par Of deeper concern is the range of evolutionary outcomes found for the same initial conditions using what are apparently an acceptable range of plausible input physics and adjustable parameters.  This is discussed next.

\par {\em Mass Limits.} An informative way to summarize a stellar evolution model is to calculate various mass limits. Examples include: The mass above which a star will form a black hole, $M_{\rm bh}$; the mass above which core helium burning ignites under non-degenerate conditions, $M_{\rm He-f}$; the mass above which an iron core forms and undergoes gravitational collapse, $M_{\rm Fe-CC}$; etc.  A summary of such limits for massive star evolution, with an emphasis on core-collapse supernova theory, is presented by \citet{heger2003}; and the fate of massive AGB stars and the potential for O-Ne-Mg core collapse are discussed by \citet{siess2007} and \citet{poelarends2008}. 

\par By comparing the mass limits found from a finely enough sampled grid of models and observational stellar data, the theoretical model can be tested. The landmark paper by \citet{maeder1981} used this approach to examine the consistency between stellar cluster data and their evolution model, and concluded that additional mixing can bring observationally inferred and theoretically calculated mass limits into better agreement, and summarized this in terms of an "overshoot" parameter $\alpha_{\rm ov} \approx 0.2$, a quantitative value which remains preferred among stellar population modelers for stars having masses larger than a few times solar \citep[e.g.][]{girardi2000, pietrinferni2004}.

\par In addition to the overshoot parameter, which is a measure of how much material mixes beyond the limits of a formally defined unstable region,  the mixing processes mentioned in \S\ref{sec:framework} also shift mass limits.  The double diffusive instability arising in thermally unstable regions which are stabilized by composition gradients, dubbed "semiconvection" \citep{merryfield1995}, is particularly important in massive star evolution. This process can change the mass of the He core following main sequence evolution, perhaps the parameter most indicative of how a massive star will end its life,  by a factor of $\sim$50\% or more \citep{woosley1988,staritsin2009}. When to include this mixing process or not, and its strengths when included, remain open questions \citep[e.g.][]{biello2001}.   

\par Fortunately, interesting constraints can be placed on stellar evolution theory by observational data when the observational uncertainties are less than a few percent.  Both asteroseismic and wide eclipsing binary data, which we discuss next, are beginning to meet these requirements.

\section{The need for precision observational data}

\begin{figure}[t]
\begin{center}
\mbox{\includegraphics[width=3.2in]{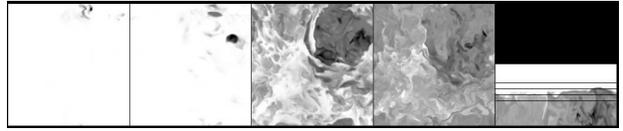}}
\end{center}
  \caption{\label{fig:entrain}  A detailed view of the boundary layer separating a turbulent convection zone from an overlying stably stratified layer in a stellar interior simulation \citep{meakin2007b,meakin2010}. The  stellar radius increases vertically upwards in the right-most panel while the other four panels are horizontal slices having radial positions indicated by the black lines in the right-most panel.  The grey scale represents oxygen abundance, with lighter values indicating higher abundance.}
\end{figure}

\par {\em Wide Eclipsing Binaries.} 
Observationally determined effective temperatures ($T_1, T_2$), luminosities ($L_1,L_2$), radii ($R_1, R_2$), and masses ($M_1,M_2$) are known for many of these systems to better than a few percent \citep{torres2010}.  In some cases precision surface rotational velocities ($[v\sin i]_1, [v\sin i]_2$) and compositions $X_{i,1} = X_{i,2}$ are also available.  A stellar evolution model can be tested against this data by comparing, e.g.,  the model radii and temperatures $(\hat{R}_1, \hat{R}_2; \hat{T}_1, \hat{T}_2) = f(M_1, M_2,X_i; t_{\rm age}, \alpha_j)$, where the hat symbol indicates model data. Here the stellar masses and compositions are also known observationally to high precision.  The age of the binary system $t_{\rm age}$ is a fitting parameter, and the theoretical model is represented by the function $f$ and the parameter set $\alpha_j$.  \citet{ribas2000} and \citet{claret2007} have studied the dependence of the convective overshoot parameter $\alpha_{\rm ov}$ on the stellar mass using this type of procedure.

\par {\em Asteroseismology.} The set of normal mode oscillations frequencies $\{\nu_k\}$ found by time monitoring stellar luminosity, together with spectroscopic ($\log g$ and $T_{\rm eff}$) and photometric information (e.g., parallax $\pi$ and luminosity $L$), provides another important test of stellar evolution theory.  In this case, the model data $\{\hat{\nu}_k,\hat{T}_{\rm eff},\log \hat{g}, \hat{L}\} = f(M,X_i,t_{\rm age};\alpha_j)$ is compared to the observed data to find a best fit stellar mass, composition\footnote{Observed values of [Fe/H] can be used to estimate the helium abundance Y by adopting a  cosmic enrichment rate.  Uncertainties and intrinsic scatter in these laws generally undermine confidence in this approach  so that Y is generally left left as a free parameter to be fit.}, age, and model parameters $\alpha_j$ \citep[see e.g.][]{vauclair2008}. The strength of this method lies in the large number of observable frequencies, each having a unique spatial dependence on internal structure \citep{unno1989}.

\par {\em Results and Implications.} A summary of the overshoot parameter and its dependence on stellar mass, inferred from wide eclipsing binary  and asteroseismological data, is presented in Fig.~\ref{fig:ov-mass}.  
While the error bars are still quite large in Fig.~\ref{fig:ov-mass}, the scatter in the data may be indicating that we are missing some essential physics.  A simple and intuitive possibility is that ``overshoot'' might not just have a mass dependence but also a time dependence. We are just beginning to collect enough precision data to investigate questions like this.  In Figure~\ref{fig:age-mass} we present a subset of wide eclipsing binary data for upper main sequence stars in the mass--age plane, showing that we can now study stars of a given mass at multiple points in time during the course of their evolution (Meakin, Sukhbold, \& Arnett, 2010 in preparation).

\section{Future directions}
\label{sec:future}

\par Precision observational data is beginning to reveal deficiencies  in our theoretical models, and improvements are needed.  More highly  parameterized models offer one means to accommodate new data, but rarely lack predictive power and even less frequently motivate new discoveries.  A more satisfying approach is to capitalize on what the new data can tell us by pursuing a more comprehensive theoretical picture.  Fortunately, many promising developments in understanding turbulent flow have already taken place in related fields of study which await application in stellar evolution \citep[see e.g.][and references therein]{canuto1992} in addition to recent work dedicated specifically to understanding stellar interior turbulence 
\citep[e.g.][]{bazan2003,browning2004,rogers2006,eggleton2006,herwig2006,meakin2006,meakin2007b,mocack2008,mocack2009,mocack2010a,mocack2010b,garaud2010}.

\par One such striking development is the recognition that a turbulent convective boundary is likely to evolve as an entrainment interface \citep[e.g.][]{fernando1991,meakin2007b,mocack2009,woodward2009}. While this synergy between geophysics and astrophysics is very intuitive and obvious in retrospect,  it nevertheless provides a more robust physical basis for describing ``overshooting'' and its time dependence, which has utility.  From relatively simple physical arguments once can deduce that the boundary of a turbulent convection zone in the deep interior should evolve at a speed given by $u = \sigma_u A Ri_{\rm B}^{-n}$ \citep{kantha1977} where $\sigma_u$ is the r.m.s.~turbulence velocity near the boundary, $Ri_{\rm B}$ is the bulk Richardson number\footnote{The bulk Richardson number is a dimensionless measure of the potential energy in stratification separating the turbulent layer from the stably stratified layer, often written in terms of the buoyancy jump $\delta b$ and integral length scale of turbulence $L$, compared to the strength of the turbulence $\sigma_u$, with $Ri_{\rm B} = \delta b L/\sigma_u^2$.  Small values indicate boundaries which are strongly distorted by the adjacent turbulence, while large values indicate relatively undisturbed boundary layers.}, $A$ is an "efficiency" parameter which depends on the turbulence in the vicinity of the interface, and $n$ is an exponent close to 1.  
The data for the stars included in Fig.~\ref{fig:age-mass} can be used to test mixing laws like the one described here, which has a natural time dependence and physical consistency built into it.  This type of theoretical modeling provides an essential bridge between turbulence simulation data, like that shown in Fig.~\ref{fig:entrain} which is used to develop these models, and reality. (See also \citet{meakin2008} and \citet{arnett2010b} for additional discussion concerning the use of simulation data to study stellar turbulence.)


\section{Conclusions}

\par Stellar evolution, particularly our understanding of massive stars, is a crucial input to many areas of astrophysical research. 
Our current best models, however, suffer severe deficiencies in treating turbulent 
transport and mixing to the degree that it is the dominant  uncertainty (in combination with 
mass loss) in our calculations. These shortcomings are currently being addressed by studying ever more realistic simulations of turbulent flow and incorporating the results from knowledge garnered in related fields of inquiry (e.g., geophysical).  This work is beginning to provide deep insights into stellar interior mixing processes.
While this work is still in a relatively formative stage, early results are beginning to reveal promising alternative approaches to the back of the envelope, mixing-length style treatments presently being used.

\par Finally, it can not be emphasized enough how important precision observational data is to testing and refining stellar evolution theory.  Fortunately, precision data is beginning to arrive just in time to begin testing the increasingly more sophisticated models of stellar interior physics that are presently being formulated.

\end{document}